\documentclass[preprint,pra,onecolumn]{revtex4-1}
\usepackage{amsmath}
\usepackage{amssymb}
\usepackage{array}
\usepackage{mathrsfs}
\usepackage{tabularx}
\usepackage{graphicx}
\usepackage{dcolumn}
\usepackage{bm}
\usepackage{graphicx}
\usepackage{color}
\usepackage{diagbox}

\begin{document}
\title{Asymmetric sending or not-sending twin-field quantum key distribution in practice}
\author{Xing-Yu Zhou$^{1,2,3}$}
\author{Chun-Hui Zhang$^{1,2,3}$}
\author{Chun-Mei Zhang$^{1,2,3}$}
\author{Qin Wang$^{1,2,3}$}
\email{qinw@njupt.edu.cn}

\affiliation{$^{1}$Institute of quantum information and technology, Nanjing University of Posts and Telecommunications, Nanjing 210003, China.}
\affiliation{$^{2}$``Broadband Wireless Communication and Senser Network Technology" Key Lab of Ministry of Education, NUPT, Nanjing 210003, China.}
\affiliation{$^{3}$``Telecommunication and Networks" National Engineering Research Center, NUPT, Nanjing 210003, China.}

\begin{abstract}
Quantum key distribution (QKD) offers a secret way to share keys between legitimate users which is guaranteed  by the law of quantum mechanics. Most recently, the limitation of transmission distance without quantum repeaters was broken through by twin-field QKD [Nature (London) \textbf{557}, 400 (2018)]. Based on its main idea, sending or not-sending (SNS) QKD protocol was proposed [Phys. Rev. A \textbf{98}, 062323 (2018)], which filled the remaining security loopholes and can tolerate large misalignment errors.  In this paper, we give a more general model for SNS QKD, where the two legitimate users, Alice and Bob, can possess asymmetric quantum channels. By applying the method present in the work, the legitimate users can achieve dramatically increased key generation rate and transmission distance compared with utilizing the original symmetric protocol. Therefore, our present work represents a further step along the progress of practical QKD.

PACS number(s): 03.67.Dd, 03.67.Hk,42.65.Lm

\end{abstract}

\maketitle

\section{Introduction}
Quantum key distribution (QKD), based on the law of quantum mechanics\cite{HK,PShor,DMayers}, allows two distant users (Alice and Bob) to establish a string of secure keys despite at the existence of the malicious eavesdropper (Eve). Since the first QKD protocol BB84 \cite{BB84} came into being, numerous protocols \cite{DecoyH,DecoyW,DecoyL,Side,MDI} were proposed to promote its development. The goal of QKD  is to own  high security and long transmission distance simultaneously. To illuminate the relationship between transmittance ($\eta $) and key rate ($R$), $R =  - \log (1 - \eta )$ is summarized \cite{PLOB} without quantum repeaters, which are regarded as a solution to overcome the limit of $R \propto O( \eta  )$. However, due to the restriction of current technology, quantum repeater is far from use \cite{QM1,QM2}. Luckily, twin-field (TF) QKD, based on the single-photon interference, with  $R \propto O(\sqrt \eta  )$ is presented \cite{TF}.

TF-QKD inherits the idea of measurement-device independent (MDI) and drastically improves the transmission distance at the same time. Upon its proposal, TF-QKD has been extensively studied \cite{TFCui,TFLin,TFWang,TFTamaki,TFYin,TFYu,TFMa,TFCurty}. Among these works, Wang \emph{et al.} developed a sending or not-sending (SNS) TF-QKD  protocol\cite{TFWang}. Without phase announcement for Z basis (signal state), SNS TF-QKD fills the remaining loophole of original TF-QKD. Moreover, due to single-photon interferences only in X basis (decoy states), SNS TF-QKD can tolerate the largest misalignment errors. This protocol seems much more practical for implementations than the original TF-QKD, and its performance has been investigated by considering statistical fluctuation and finite numbers of phase slices \cite{TFYu}.

In real life, most locations of users are not on the symmetry of untrusted third party (UTP). Especially in a multi-user network, UTP can hardly locate at the centre of all users. One simple solution is to add extra fibers or attenuations at the closer side to compensate the difference between the two transmittances, where original symmetric protocol is certainly suitable. This seems to be a "$buckets \; effect$", and the final key rate is limited by the smaller transmittance.

In this work, we develop a general model for the SNS TF-QKD, where the two parties possess asymmetric quantum channels.
  Different from previous works on asymmetric MDI-QKD \cite{AsyXu,AsyHu,AsyWang}, decoy-state method can not applied directly in asymmetric SNS TF-QKD. According to our analysis, decoy-state method still can be an efficient and secure method in present work only by satisfying some extra constraints.

The paper is organized as follow: In Sec \ref{1}, we will introduce some basic steps on how to implement asymmetric SNS TF-QKD. Besides, decoy-state method and other theoretical models are given. Corresponding numerical simulations are shown in Sec \ref{3}. Finally, summaries and outlooks are given out in Sec \ref{4}.

\section{The Decoy-State Asymmetric SNS TF-QKD}\label{1}
In this section, without adding extra compensation of transmittance, we will show the possibility of applying  decoy state asymmetric SNS TF-QKD only by adjusting dependent intensities and other parameters.

 \subsection{Basic steps of decoy-state asymmetric SNS TF-QKD}
\begin{figure}[htbp]
\centering
\includegraphics[width=12cm]{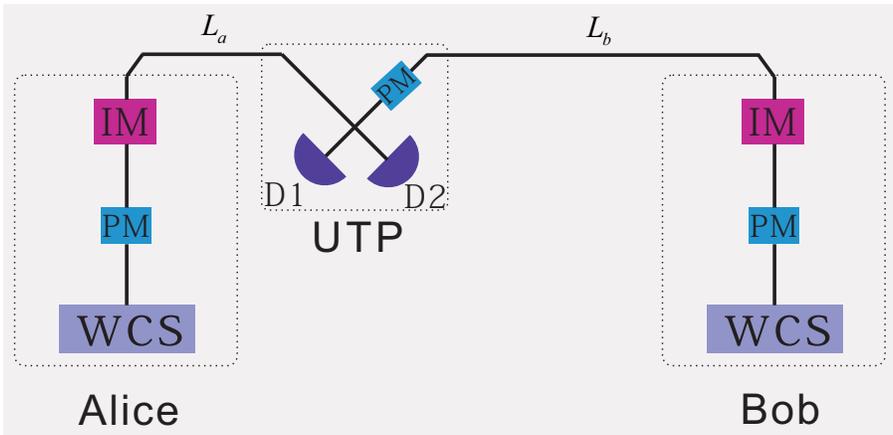}
\caption{Schematic setup of asymmetric SNS TF-QKD. WCS: weak coherent source; PM: phase modulator; IM: intensity modulator; D1(D2): single-photon detector.  Bob is farther from UTP than Alice; $L_a$ and $L_b$ are the distance between user and the UTP respectively.}
\label{F1}
\end{figure}

Below let us describe the detailed SNS TF-QKD in asymmetric situation. Corresponding schematic setup is shown in Fig.\ref{F1}.

(0) For each time window, $i$, Alice and Bob send a strong reference light with coherent state pulses to the UTP.  Besides, they add  extra random phases ${\delta _{a}}$ and ${\delta _{b}}$ to their pulses. Here, we denote the distance between user and the UTP as $L_a$ and $L_b$ ($L_a < L_b$ ) respectively.

(1) Alice (Bob) randomly chooses signal window (Z-window) and decoy  window (X-window) with probability ${P_{za}}$ (${P_{zb}}$) and ${1-P_{za}}$ (${1-P_{zb}}$). In Z-window, Alice (Bob) determines to send a signal state pulse $\left| {\sqrt {{u_a  }} {e^{i{\delta _a} + i{\gamma _a}}}} \right\rangle $ ($\left| {\sqrt {{u_b}} {e^{i{\delta _b} + i{\gamma _b}}}} \right\rangle $) with probability ${\varepsilon _a}$ (${\varepsilon _b}$), and not to send with $1 - {\varepsilon _a}$ $(1 - {\varepsilon _b})$; In X-window, Alice and Bob emit decoy state pulse $\left| {\sqrt {\alpha } {e^{i{\delta _a} + i{\gamma _a}}}} \right\rangle $and $\left| {\sqrt {{\beta }} {e^{i{\delta _b} + i{\gamma _b}}}} \right\rangle $, respectively. ${{\alpha }} \in \{ {v_a},{w_a},o\} $; ${{\beta }} \in \{ {v_b},{w_b},o\} $.
 ${\gamma _a}$ and ${\gamma _b}$ are the global phase.  Note that, in asymmetric situation, Alice is reasonably assumed to be closer to the UTP than Bob. Then, she should postpone the emission for $\tau $ time-windows to ensure the synchronization, i.e, the two states chosen at the same time-windows reach the beam splitter simultaneously.

(2) After the UTP performs  the phase compensation  with the aid of strong reference light,  the two-mode state turns into, for example, $\left| {\sqrt {{\alpha}} {e^{i{\delta _a}}}} \right\rangle \left| {\sqrt {{\beta}} {e^{i{\delta _b}}}} \right\rangle $. Then the UTP measures the incoming pulses and records the clicking or non-clicking events of the two detectors.

(3)The UTP announces the measurement outcomes after the distribution progress ends. Then, the users announce for each pulse  whether it is a Z-window or an X-window.  The intensity and extra phase of X-window should also be public.
An efficient event is defined as the following two cases: (a)  Alice and Bob both choose Z-windows and only one detector clicking at UTP's side. In this case, four events and the corresponding raw keys are shown in Table \ref{T1};
(b) Alice and Bob both choose the corresponding intensities in X-window when UTP announces single clicking of detectors, and the  phases ${\delta _a}$ , ${\delta _b}$  should satisfy either of following two inequations:
\begin{align}\label{distribution}
\left| {{\delta _a} - {\delta _b} } \right| \le \frac{{2\pi }}{M} ,  \qquad   \left| {{\delta _a} - {\delta _b} - \pi } \right| \le \frac{{2\pi }}{M}.
\end{align}
$M$ is the number of phase slices pre-determine by users.

\begin{table}[!htbp]
\centering\caption{When an effective event happened in Z-window, if Alice (Bob) decides  to send a signal pulse, she (he) records a bit 1 (0); if  Alice (Bob) decides not to send a signal pulse, she (he) records a bit 0 (1); }
\begin{tabular}{|c|c|c|}
\hline
\diagbox{Alice}{keys}{Bob}&Sending&Not-sending\\
\hline
Sending&10&11\\
\hline
Not-sending&00&01\\
\hline
\end{tabular}
\label{T1}
\end{table}

(4)After the announcement, Alice and Bob get the gain in Z-window. By sacrificing some bits in Z-window, Alice and Bob get the average quantum bit error rate (QBER) in Z basis. Besides, they estimate the single-photon yield $Y_1$ and the error rate $e_1$ by observed values in X-window.

(5)Error correction and privacy amplification are performed before calculating the final secret keys.

 \subsection{Decoy-state method and theoretical models}
Before introducing the decoy-state formulae of this protocol, we will review its essence. In a decoy-state method \cite{DecoyH,DecoyW,DecoyL}, legitimate users need to modulate light pulses into different intensities and post-announce the details. Eve can not distinguish which one is the signal state pulse, and can only carry out identical attacking strategies in quantum channels. As a result, the photon-number-splitting attacks will affect the yields, $Y_n$, and QBER, $e_n$ which only depend on the numbers of photons $n$.  Whether an eavesdropper exists can be judged from the reasonability of  $Y_n$ and  $e_n$.  In essence, decoy state method is based on the following equations:
\begin{align}
{Y_n}(signal) = {Y_n}(decoy), \nonumber \\
{e_n}(signal) = {e_n}(decoy).
\end{align}

Now let us come to the asymmetric SNS TF-QKD: Denote Alice and Bob send pulses with intensities $x_a,x_b$ respectively, and corresponding transmittances are ${\eta _a},{\eta _b}$ $({\eta _a} > {\eta _b})$. For simplicity, we assume that the two detectors at UTP's sides are identical and each with a dark count rate $P_d$ and detection efficiency  ${\eta _d}$ individually. 

The counting rate of the n-photon states which causes effective events can be written as:
\begin{align}\label{2222}
Q_n^{{x_a}{x_b}} = \sum\limits_{m = 0}^n {\frac{{{e^{ - {x_a}}}{x_a}^m}}{{m!}}} \frac{{{e^{ - {x_b}}}{x_b}^{n - m}}}{{(n - m)!}}[1 - {(1 - {P_d})^2}{(1 - {\eta _a})^m}{(1 - {\eta _b})^{n - m}}] .
\end{align}
Hereafter, we call the above event as the n-photon effective event.
Considering it may possess $m$ photons from Alice and ($n-m$) photons form Bob, the equivalent photon number distribution can be formulated as: 
\begin{align}\label{111}
{P_n}({x_a} + {x_b}) = \sum\limits_{m = 0}^n {\frac{{{e^{ - {x_a}}}{x_a}^m}}{{m!}}} \frac{{{e^{ - {x_b}}}{x_b}^{n - m}}}{{(n - m)!}} .
\end{align}

Correspondingly, the equivalent yield of the n-photon effective event can be expressed as:
\begin{align}\label{333}
Y_n^{{x_a}{x_b}} =& \frac{{Q_n^{{x_a}{x_b}}}}{{{P_n}({x_a} + {x_b})}}\nonumber \\
=& 1 - {(1 - {P_d})^2}{[\frac{{{x_a}(1 - {\eta _a}) + {x_b}(1 - {\eta _b})}}{{{x_a} + {x_b}}}]^n} \nonumber \\
=& 1 - {(1 - {P_d})^2}{[\frac{{k(1 - {\eta _a}) + (1 - {\eta _b})}}{{k + 1}}]^n},
\end{align}
where $k=\frac{{{x_a}}}{{{x_b}}}$.
Obviously, in the asymmetric case, the value of  $Y_n^{{x_a}{x_b}}$ is not only dependent on the photon numbers ($n$), but also related to the ratio ($k$) of two intensities. Therefore, the original lower bound of the single-photon counting rate ($Y_1$) and upper bound of the single-photon error rate ($e_1$) cannot be applied directly. In the \textbf{Appendix}, we will give corresponding proof for the renewed formulae.    

In Eq.(\ref{333}),  $Y_n^{{x_a}{x_b}}$ is concerned with the ratio $k$. For convenience, we denote $Y_n^{{x_a}{x_b}}$ ($Q_n^{{x_a}{x_b}}$)as $Y_n^{k}$ ($Q_n^{k}$);
Denote ${w_a} + {w_b} = {\mu _1}, {v_a} + {v_b} = {\mu _2}, \frac{{{w_a}}}{{{w_b}}} = {k_1},\frac{{{v_a}}}{{{v_b}}} = {k_2}.$
According to the analysis in the \textbf{Appendix}, for  ${k_1}  \le  {k_2}$,  we can get the lower bound of single-photon yield in X-window 
\begin{equation}\label{Y}
Y_1^L=\frac{{{P_2}({\mu _2}){Q_{{\mu _1}}} - {P_2}({\mu _1}){Q_{{\mu _2}}} + [{P_2}({\mu _1}){P_0}({\mu _2}) - {P_2}({\mu _2}){P_0}({\mu _1})]{Y_0}}}{{{P_2}({\mu _2}){P_1}({\mu _1}) - {P_2}({\mu _1}){P_1}({\mu _2})}}.
\end{equation}

In addition, to estimate the single-photon yield in Z-window, a restriction on the ratio of intensities, e.g., $\frac{{{u_1}}}{{{u_2}}} \ge  {k_1}$, should be imposed. In this case, the yield in X-window is not larger than the yield in Z-window. Thus, $Y_1^L$ can also be looked as the lower bound in Z-window.
Accordingly, the QBER of single-photon pulses is given by \cite{TFWang}:
\begin{equation}\label{e}
{e_1} \le e_1^U = \frac{{{Q_{{\mu _1}}}{E_{{\mu _1}}} - {P_0}({\mu _1}){Y_0}{e_0}}}{{{P_1}({\mu _1}){Y_1^L}}},
\end{equation}
where ${e_0}=0.5$.

In real-life implementations, the average counting rate and QBER in X-window can be directly measured. In this work, we use a linear model to predict what it should be observed in experiment. Consider a two-mode state $\left| {\sqrt {{\alpha}} {e^{i{\delta _a}}}} \right\rangle \left| {\sqrt {{\beta}} {e^{i{\delta _b}}}} \right\rangle $ goes through the quantum channels and a
beam-splitter. It turns into $\left| {\sqrt {\frac{{\alpha {\eta _a}}}{2}} {e^{i{\delta _a}}} + \sqrt {\frac{{\beta {\eta _b}}}{2}} {e^{i{\delta _b}}}} \right\rangle  \otimes \left| {\sqrt {\frac{{\alpha {\eta _a}}}{2}} {e^{i{\delta _a}}} - \sqrt {\frac{{\beta {\eta _b}}}{2}} {e^{i{\delta _b}}}} \right\rangle $. The corresponding gains ($Q_{\alpha \beta }^{{\delta _a}{\delta _b}}$) and the quantum-bit errors ($Q_{\alpha \beta }^{{\delta _a}{\delta _b}}E_{\alpha \beta }^{{\delta _a}{\delta _b}}$) are given by
\begin{equation}\label{QQQ}
Q_{\alpha \beta }^{{\delta _a}{\delta _b}}
= (1 - {P_d}){e^{ - \frac{{\alpha {\eta _a}}}{2} - \frac{{\beta {\eta _b}}}{2}}}({e^{ - \cos ({\delta _a} - {\delta _b})\sqrt {\alpha \beta {\eta _a}{\eta _b}} }} + {e^{\cos ({\delta _a} - {\delta _b})\sqrt {\alpha \beta {\eta _a}{\eta _b}} }}) - 2{(1 - {P_d})^2}{e^{ - \alpha {\eta _a} - \beta {\eta _b}}},
\end{equation}
\begin{equation}\label{QEQEQE}
Q_{\alpha \beta }^{{\delta _a}{\delta _b}}E_{\alpha \beta }^{{\delta _a}{\delta _b}} = (1 - {P_d}){e^{ - \frac{{\alpha {\eta _a}}}{2} - \frac{{\beta {\eta _b}}}{2} - \cos ({\delta _a} - {\delta _b})\sqrt {\alpha \beta {\eta _a}{\eta _b}} }} - {(1 - {P_d})^2}{e^{ - \alpha {\eta _a} - \beta {\eta _b}}}.
\end{equation}
After phase post-selection in X-window, $|{\delta _a} - {\delta _b}|$ are ranging among $[0,\frac{{2\pi }}{M}] \cup [\pi ,\pi  + \frac{{2\pi }}{M}]$. Define the system error rate as ${E_{sys}} = \frac{1}{2} - \frac{{\sqrt {{x_1}{\eta _1}{x_2}{\eta _2}} }}{{{x_1}{\eta _1} + {x_2}{\eta _2}}} + \frac{{2\sqrt {{x_1}{\eta _1}{x_2}{\eta _2}} }}{{{x_1}{\eta _1} + {x_2}{\eta _2}}}{E_d}$, where ${E_d}$ is the build-in misalignment error of the optical system. Here ${E_{sys}}$ comes from single-photon interference and leads to an extra equivalent phase differences between Alice and Bob, denoted as $\Delta  = \arccos (1 - 2{E_{sys}})$. By integrating, the average gain and quantum-bit errors can be expressed as
\begin{equation}\label{Q}
{Q_{\alpha \beta }}=\frac{{{M^2}}}{{4{\pi ^2}}}\int_\Delta ^{\frac{{2\pi }}{M} + \Delta } {\int_0^{\frac{{2\pi }}{M}} {Q_{\alpha \beta }^{{\delta _a}{\delta _b}}d{\delta _a}} } d{\delta _b},
\end{equation}
 \begin{equation}\label{QE}
{Q_{\alpha \beta }}{E_{\alpha \beta }} = \frac{{{M^2}}}{{4{\pi ^2}}}\int_\Delta ^{\frac{{2\pi }}{M} + \Delta } {\int_0^{\frac{{2\pi }}{M}} {Q_{\alpha \beta }^{{\delta _a}{\delta _b}}E_{\alpha \beta }^{{\delta _a}{\delta _b}}d{\delta _a}} } d{\delta _b}.
 \end{equation}


Finally, with the above formulae, the key generation rate can be expressed as
\begin{equation}
R = {P_{za}}{P_{zb}}\{ [{\varepsilon _a}(1 - {\varepsilon _b}){e^{ - {u_a}}}{u_a} + {\varepsilon _b}(1 - {\varepsilon _a}){e^{ - {u_b}}}{u_b}]Y_1^L[1 - H(e_1^U)] - {Q_{{u_a}{u_b}}}fH({E_{{u_a}{u_b}}})\},
\end{equation}
where ${Q_{{u_a}{u_b}}}$ and ${E_{{u_a}{u_b}}}$ are the average gain and QBER of effective events in Z-window;$f$ is the error correction efficiency; $H(\xi ) =  - \xi {\log _2}(\xi ) - (1 - \xi ){\log _2}(1 - \xi )$.
\section{Numerical simulations}\label{3}

With all the above formulae, we can now carry out numerical simulations for the asymmetric SNS TF-QKD. To be noted, for the asymmetric case, in order to reach the highest visibility of single-photon interference in the UTP's side, certain constraints should be set on the system parameters, to make light from each path possessing the same intensity before the beam-splitter. Statistical fluctuation is also taken into account. For simplicity, we make a Gaussian distribution assumption of the channel fluctuations and apply the standard deviation method, setting the failure probability as $10^{-7}$ \cite{yu2}.
Finite number of phase slices, $M$, is considered.
The experimental parameters used here are taken from Ref.\cite{TFYu}, which are listed out in Table \ref{T2}. Besides, global optimization is applied for a better performance.

\begin{center}
\begin{table}[hbp]
  \caption{Parameters for simulations. $\eta_d$ and $P_d$ are detection efficiency and dark counting rate per pulse of  UTP's detectors respectively; $E_d$ is misalignment error of optical system; $ \alpha$ is the transmission fiber loss constant; $f$ is the error correction efficiency; $N$ is the number of total pluses.}
\renewcommand{\arraystretch}{1.3}
\begin{tabularx}{\linewidth}{XXXXXX}  \hline \hline
    $\eta_d$ & $P_d$&  $E_d$ &    $ \alpha (dB/km ) $  & $f$ & $N$\\ \hline
    50\%& $1\times10^{-10}$  & 15\%   &  0.2 &1.1 &$10^{12}$\\ \hline\hline
  \end{tabularx}
 \label{T2}
\end{table}
\end{center}

\begin{figure}[ptb]
\begin{center}
\includegraphics[scale=0.3]{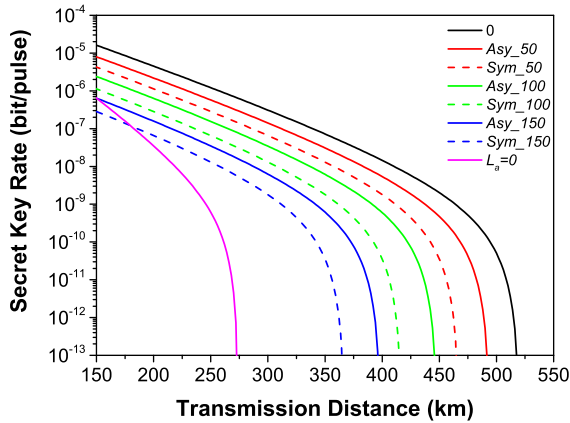}
\end{center}
\caption{Result of secret key rate with respect to the total transmission distance ($L_a+L_b$). The line marked with '0' represents the original symmetric case;  '$La=0$' represents the special case that Alice and UPT are together. Among the rest lines, Label '$Asy$' means the asymmetric method in this work; '$Sym$' is for directly using the symmetric method by adding extra attenuations; The numbers in the label are the value of ${L_b}-{L_a}$.  }
\label{Fig2}
\end{figure}

In Fig. \ref{Fig2}, $L_a (L_b)$ is the distance between Alice (Bob) and the UTP. As mentioned above, by adding extra attenuations,  QKD system with asymmetric channels can be transformed into a symmetric one. Hereafter, we call it the original symmetric method. For a vivid comparison, we plot the secret key rate by using two different methods: the original symmetric method  and the asymmetric method proposed in this work. Obviously, the present asymmetric work significantly improves both the secret key rate and the transmission distance compared with the  original symmetric one. Consider an extreme case, where UTP and Alice are in the same lab ($L_a=0$). It seems like a BB84 protocol with two parties and the transmission distance is about half of the symmetric case.  By analogy, we can regard the key rate of asymmetric SNS TF-QKD and transmittance as a relationship of $R \propto O({\eta ^\sigma  })$, $\sigma   \in (\frac{1}{2},1)$.

\begin{figure}[ptb]
\begin{center}
\includegraphics[scale=0.3]{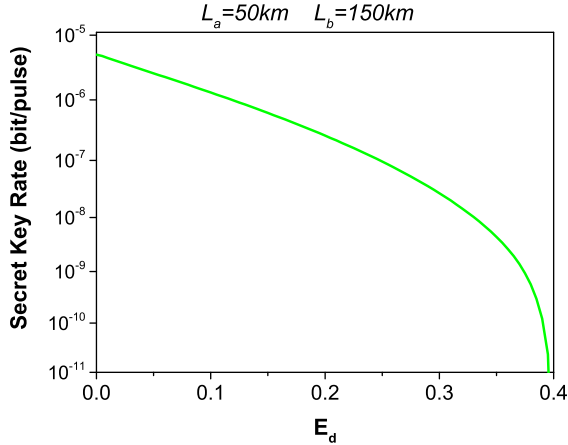}
\end{center}
\caption{Secret key rate as a function of the $E_d$ when $L_a$=50km, $L_b$=150km.}
\label{Fig3}
\end{figure}

Moreover, as we can see from Fig.\ref{Fig3}, in an asymmetric case, the SNS TF-QKD can tolerate very high misalignment errors, e.g., it can still generate secret keys even when $E_d$ exceeds 0.35. In original TF-QKD, the single-photon interference is a challenging technology and may cause large misalignment errors. While in the SNS TF-QKD, it only needs sending or-not sending pulses instead of interference in Z-window, and can thus tolerate much larger misalignment errors. In the present asymmetric case, the advantages hold on and make it very promising candidate in practical applications.

\section{Summaries and outlooks}\label {4}
In conclusion, we have extended the SNS TF-QKD to the asymmetric case and given a general model. The lower bound of the yield and the upper bound of the QBER for single-photon contributions have been rederived. Then the asymmetry of channels could be compensated by optimizing the adjustable system parameters.  Through implementing full parameter optimization on the numerical simulations, we demonstrate that our new method can dramatically improve the key generation rate and the transmission distance compared with the original symmetric method. In addition, some state-of-the-art optimization techniques, such as collective constraints and joint estimations can be applied to further improve the performance  \cite{TFYu,yu3}. Therefore, our work represent a further step towards practical application of the QKD.

We gratefully acknowledge the financial support from the National Key R$\&$D Program of China (Grant Nos. 2018YFA0306400, 2017YFA0304100), the National Natural Science Foundation of China (Grants Nos. 61475197, 61590932, 11774180, 61705110, 11847215), the China Postdoctoral Science Foundation (Grant No. 2018M642281), the Natural Science Foundation of Jiangsu Province (Grant No. BK20170902), the Jiangsu Planned Projects for Postdoctoral Research Funds(Grant No. 2018K185C), and the Postgraduate Research and Practice Innovation Program of Jiangsu Province (Grant No. KYCX17$\_$0791).

\section*{Appendix}
Below, we will give a detailed derivation of Eq. (\ref{Y}). First, to get the monotonicity of $Y_n^{k}$, we make a formula deformation:
\begin{align}
Y_n^{k}{\rm{ = }}1 - {(1 - {P_d})^2}{[\frac{{k(1 - {\eta _a}) + (1 - {\eta _b})}}{{k + 1}}]^n} \nonumber\\
 = 1 - {(1 - {P_d})^2}{[(1 - {\eta _a}) + \frac{{{\eta _a} - {\eta _b}}}{{k + 1}}]^n}.
\end{align}
Obviously, when ${\eta _a} > {\eta _b}$, $Y_n^{k}$ is an increasing function of $k$.

 The average gains of the two decoy-states is given by:
\begin{equation}\label{Qmu1}
{Q_{{\mu _1}}} = {Y_0}{P_0}({\mu _1}) + Y_1^{{k_1}}{P_1}({\mu _1}) + Y_2^{{k_1}}{P_2}({\mu _1}) + \sum\limits_{n = 3}^\infty  {Y_n^{{k_1}}{P_n}({\mu _1})},
\end{equation}
\begin{equation}\label{Qmu2}
{Q_{{\mu _2}}} = {Y_0}{P_0}({\mu _2}) + Y_1^{{k_2}}{P_1}({\mu _2}) + Y_2^{{k_2}}{P_2}({\mu _2}) + \sum\limits_{n = 3}^\infty  {Y_n^{{k_2}}{P_n}({\mu _2})}.
\end{equation}

When ${k_1}\le{k_2}$, $Y_n^{k_1} \le Y_n^{k_2}$ holds on.
Eq.(\ref{Qmu2}) can be expressed as:
\begin{equation}
{Q_{{\mu _2}}} = {Y_0}{P_0}({\mu _2}) + Y_1^{{k_1}}{P_1}({\mu _2}) + Y_2^{{k_1}}{P_2}({\mu _2}) + \sum\limits_{n = 3}^\infty  {Y_n^{{k_1}}{P_n}({\mu _2})}  + {\Delta _1},
\end{equation}
where ${\Delta _1} = (Y_1^{{k_2}} - Y_1^{{k_1}}){P_1}({\mu _2}) + (Y_2^{{k_2}} - Y_2^{{k_1}}){P_2}({\mu _2}) + \sum\limits_{n = 3}^\infty  {(Y_n^{{k_2}} - Y_n^{{k_1}}){P_n}({\mu _2})}  \ge 0$.

By using the similar method as in Ref. \cite{yu2}:
\begin{equation}\label{Qmu133}
{P_2}({\mu _2}){Q_{{\mu _1}}} = {P_2}({\mu _2}){Y_0}{P_0}({\mu _1}) + {P_2}({\mu _2})Y_1^{{k_1}}{P_1}({\mu _1}) + {P_2}({\mu _2})Y_2^{{k_1}}{P_2}({\mu _1}) + {P_2}({\mu _2})\sum\limits_{n = 3}^\infty  {Y_n^{{k_1}}{P_n}({\mu _1})},
\end{equation}
 \begin{equation}\label{Qmu144}
{P_2}({\mu _1}){Q_{{\mu _2}}} = {P_2}({\mu _1}){Y_0}{P_0}({\mu _2}) + {P_2}({\mu _1})Y_1^{{k_1}}{P_1}({\mu _2}) + {P_2}({\mu _1})Y_2^{{k_1}}{P_2}({\mu _2}) + {P_2}({\mu _1})\sum\limits_{n = 3}^\infty  {Y_n^{k1}{P_n}({\mu _2})}  + {P_2}({\mu _1}){\Delta _1}.
 \end{equation}

Combining Eq.(\ref{Qmu133}) and Eq.(\ref{Qmu144}), we can get:
\begin{align}
&{P_2}({\mu _1}){Q_{{\mu _2}}} - {P_2}({\mu _2}){Q_{{\mu _1}}} \nonumber\\
=&[{P_2}({\mu _1}){P_0}({\mu _2}) - {P_2}({\mu _2}){P_0}({\mu _1})]{Y_0} + [{P_2}({\mu _1}){P_1}({\mu _2}) - {P_2}({\mu _2}){P_1}({\mu _1})]Y_1^{{k_1}}   \nonumber\\
&+ {P_2}({\mu _1})\sum\limits_{n = 3}^\infty  {Y_n^{{k_1}}{P_n}({\mu _2})}  - {P_2}({\mu _2})\sum\limits_{n = 3}^\infty  {Y_n^{{k_1}}{P_n}({\mu _1})} + {P_2}({\mu _1}){\Delta _1}.
\end{align}

Denote ${P_2}({\mu _1})\sum\limits_{n = 3}^\infty  {Y_n^{{k_1}}{P_n}({\mu _2})}  - {P_2}({\mu _2})\sum\limits_{n = 3}^\infty  {Y_n^{{k_1}}{P_n}({\mu _1})}  = {\Delta _2}$. Due to the weak coherent source satisfying the following condition \cite{yu2}:
\begin{equation}\label{condition1}
\frac{{{P_n}({\mu _2})}}{{{P_n}({\mu _1})}} \ge \frac{{{P_2}({\mu _2})}}{{{P_2}({\mu _1})}} \ge \frac{{{P_1}({\mu _2})}}{{{P_1}({\mu _1})}},
\end{equation}
we can conclude that ${\Delta _2} > 0$.

Finally, when ${k_1}  \le  {k_2}$, the lower bound of the single-photon yield:
\begin{equation}\label{YYYYYY}
Y_1^{{k_1}} \ge Y_1^L = \frac{{{P_2}({\mu _2}){Q_{{\mu _1}}} - {P_2}({\mu _1}){Q_{{\mu _2}}} + [{P_2}({\mu _1}){P_0}({\mu _2}) - {P_2}({\mu _2}){P_0}({\mu _1})]{Y_0}}}{{{P_2}({\mu _2}){P_1}({\mu _1}) - {P_2}({\mu _1}){P_1}({\mu _2})}}.
\end{equation}

\end{document}